\documentclass[final,5p,times,twocolumn]{elsarticle}

\usepackage{amssymb}
\usepackage{multirow}
\usepackage{subfigure}
\newcommand{\bma}[1]{\mbox{\boldmath${#1}\/$}}
\newcommand{\Cdot}{\bma{\cdot}}
\newcommand{\Nabla}{\bma{\nabla}}

\journal{Aerospace Science and Technology}

\begin{document}

\begin{frontmatter}

\title{Design of the magnetic diagnostics unit onboard LISA Pathfinder}

\author[FA,IEEC]{Marc~Diaz-Aguil\'o}
\author[ICE,IEEC]{Ignacio~Mateos}
\author[UPC]{Juan~Ramos-Castro}
\author[ICE,IEEC]{Alberto~Lobo}
\author[FA,IEEC]{Enrique~Garc\'\i a--Berro}

\address[FA]{Departament de F\'\i sica  Aplicada, 
             Universitat Polit\`ecnica  de Catalunya,
             c/Esteve Terrades 5, 
             08860 Castelldefels, 
             Spain}
\address[UPC]{Departament d'Enginyeria Electr\`onica,
              Universitat Polit\`ecnica  de Catalunya,
              c/Jordi Girona 1-3,
              08034 Barcelona,
              Spain}
\address[ICE]{Institut de Ci\`encies de l'Espai, CSIC,
              Campus UAB, Facultat  de   Ci\`encies,
              Torre C-5,
              08193 Bellaterra, 
              Spain}
\address[IEEC]{Institut d'Estudis Espacials de Catalunya,  
               c/Gran Capit\`a 2-4, 
               Edif. Nexus 104,
               08034 Barcelona, 
               Spain}

\begin{abstract}
LISA (Laser  Interferometer Space Antenna)  is a joint mission  of ESA
and NASA  which aims  to be the  first space-borne  gravitational wave
observatory. Due  to the high complexity  and technological challenges
that  LISA   will  face,  ESA   decided  to  launch   a  technological
demonstrator, LISA Pathfinder.  The  payload of LISA Pathfinder is the
so-called LISA Technology Package, and will be the highest sensitivity
geodesic  explorer flown  to  date.  The  LISA  Technology Package  is
designed to measure relative  accelerations between two test masses in
nominal  free  fall  (geodesic  motion).  The  magnetic,  thermal  and
radiation disturbances  affecting the payload are  monitored and dealt
by the  diagnostics subsystem.  The diagnostics  subsystem consists of
several modules,  and one of  these is the magnetic  diagnostics unit.
Its main function is the assessment of differential acceleration noise
between test masses  due to the magnetic effects. To do  so, it has to
determine  the magnetic  characteristics  of the  test masses,  namely
their magnetic remanences and susceptibilities.  In this paper we show
how this can be achieved to the desired accuracy.
\end{abstract}


\end{frontmatter}

\section{Introduction}
\label{sec:Introduction}

LISA (Laser Interferometer  Space Antenna) is a space  mission of NASA
and  ESA which aims  at detecting  low frequency  gravitational waves.
LISA will consist in a constellation of three spacecraft occupying the
vertexes of an equilateral triangle of side 5 million kilometers.  The
barycenter of the constellation will  orbit around the Sun following a
quasi-circular orbit inclined $1^\circ$  with respect to the ecliptic,
and trailing  the Earth  by some $20^\circ$.   Each of  the spacecraft
harbors  two  proof   masses,  carefully  protected  against  external
disturbances such  as solar radiation pressure  and charged particles,
which  ensures they  are in  nominal free-fall  in  the interplanetary
gravitational  field.   Gravitational waves  show  up as  differential
accelerations between pairs of proof  masses, and the main aim of LISA
is  to  measure such  acceleration  using  laser interferometry.   The
interested  reader is referred  to~\cite{bib:axel} and~\cite{bib:bill}
for more extensive information.

\begin{figure}[t]
\centering
\includegraphics[width=0.9\columnwidth]{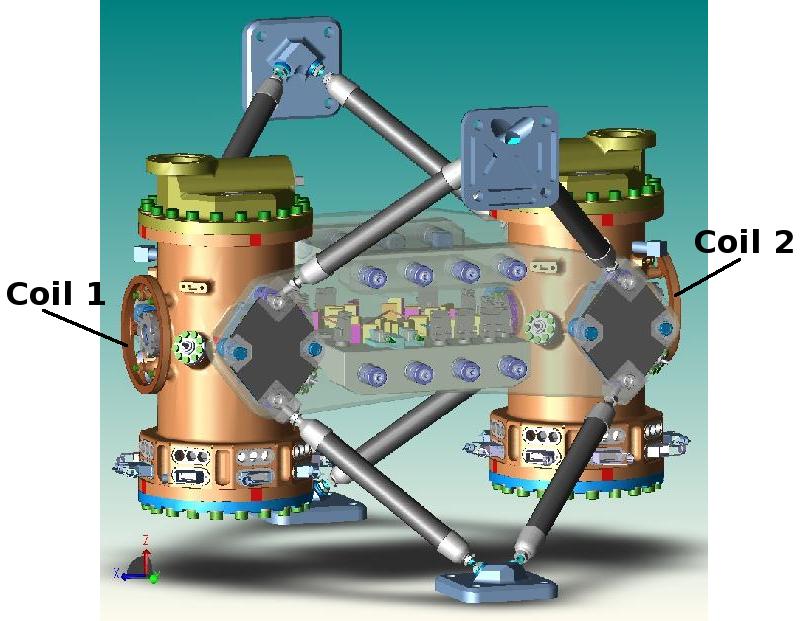}
\caption{Artistic view of the  LISA Technology Package. The two towers
         are the Gravitational  Reference Sensors.  They are connected
         by  the optical  bench  (grey), where  the interferometer  is
         located.  The induction  coils are  located next  to  the two
         towers.}
\label{fig.LTP}
\end{figure}

The  technologies   required  for  the  LISA  mission   are  many  and
challenging.  This,  coupled with the  fact that some  flight hardware
cannot be  tested on  ground, led ESA  to frame within  its Scientific
Program  a  technology  demonstrator  to test  the  required  critical
technologies in a flight  environment.  Its launch is expected towards
early 2014.  The  basic goal of LISA Pathfinder  consists in measuring
to the  highest possible accuracy  the acceleration noise of  two test
masses 35  centimeters away,  and to keep  this noise below  a certain
limit~\cite{bib:lisaPF,  bib:LPFmission}.  The  payload on  board LISA
Pathfinder is called the LISA Technology Package (LTP)~\cite{bib:anza,
  bib:DDS_LTP}.   Its  main   components  are  the  two  Gravitational
Reference Sensors (GRSs) --- shown as the two large vertical cylinders
in figure~\ref{fig.LTP} --- and the Optical Metrology System (OMS) ---
which is  an interferometer placed on the  horizontal platform between
them. The GRS  consists in a set of electrodes  aimed to determine the
position of  a test mass with  respect to the  spacecraft to nanometer
precision, using capacitance  measurements. The OMS provides picometer
precision measurements  of the relative  position of two  test masses.
Using these measurements, a  set of micro-thrusters --- Field Emission
Electric  Propulsion (FEEP) ---  relocate the  spacecraft so  that the
test   mass   preserves   a   free   fall   motion.    In   the   same
figure~\ref{fig.LTP},  the  location of  two  induction  coils can  be
observed.  The  coils will be  used to generate a  controlled magnetic
field within the  volume of the LTP Core Assembly (LCA)  in such a way
that  the   magnetic  characteristics  of  the  test   masses  can  be
determined.  At the same time,  the magnetic field inside the LCA will
be measured  by a  set of four  tri-axial fluxgate  magnetometers (not
shown in  the figure).   All these pieces  of hardware  constitute the
Magnetic  Diagnostic  Subsystem.  Finally,  the  Data Management  Unit
(DMU) is  in charge of commanding  and acquiring signals  from all the
mentioned subsystems.

Magnetic  noise in  the LTP  can be  a significant  part of  the total
readout  noise, 1.2$\cdot$10$^{-14}$\,m\,s$^{-2}$\,Hz$^{-1/2}$  out of
3.0$\cdot$10$^{-14}$\,m\,s$^{-2}$\,Hz$^{-1/2}$    is   the   allocated
maximum acceleration magnetic noise budget.  This noise occurs because
the  residual magnetization  and  susceptibility of  the proof  masses
couple  with a  surrounding magnetic  field,  giving rise  to a  force
\cite{bib:jackson}:

\begin{equation}
 {\bf F} = \left\langle\left[\left({\bf M} + 
           \frac{\chi}{\mu_0}\,{\bf B}\right)\Cdot\Nabla\right]{\bf B}
           \right\rangle V\, ,
 \label{eq.1}
\end{equation}

\noindent and a torque:

\begin{equation}
 \textbf{N} = \left\langle\textbf{M}\times\textbf{B} +
 \textbf{r}\times \left[\left(\textbf{M}\Cdot\Nabla\right)\,\textbf{B}\right]
 \right\rangle V
 \label{eq.1n}
\end{equation}

\noindent In  these expressions {\bf B}  is the magnetic  field in the
test mass, {\bf M} stands for  the magnetization of the test mass, $V$
is the volume of the test mass, $\chi$ is its magnetic susceptibility,
$\mu_0$ is  the vacuum magnetic  constant, and \textbf{r}  denotes the
distance  to the  center  of  the test  mass.   Finally, the  notation
$\langle\cdots\rangle$ refers  to the  volume average of  the enclosed
quantity. Thus,  to estimate and ultimately  subtract the acceleration
noise due to the magnetic interactions, the magnetic properties of the
test masses must be determined.

In this paper we describe the experimental setup and the data analysis
needed  to infer the  values of  the magnetic  properties of  the test
masses on  board the LTP, and  we assess the  feasibility of obtaining
the magnetic  characteristics of the  test masses with  good accuracy.
Specifically, we present a set  of simulations aimed at evaluating the
response  of the  LTP hardware  (coils  and test  masses) and  control
architecture  (drag  free  controllers  and low  frequency  suspension
controllers)  when a controlled  magnetic field  is applied.   We will
show  that  using  this  procedure,  both the  magnetization  and  the
magnetic susceptibility of  the proof masses can be  determined to the
desired  accuracy.  The  paper is  organized as  follows.   In section
\ref{sec:Experiment},   we   describe  the   main   elements  of   the
experimental  setup.  Section \ref{sec:forces}  is devoted  to compute
the forces and  torques acting on the test  masses. It follows section
\ref{sec:NoiseSources}, where  the different noise  sources perturbing
the experiment  are presented.  Finally,  in section \ref{sec:Results}
we present  our results,  whereas in section  \ref{sec:Conclusions} we
summarize our main findings and we present our conclusions.

\section{Experiment description}
\label{sec:Experiment}

As mentioned, the basic  approach to determine the magnetic properties
of the test  masses is to inject a controlled  signal with the onboard
coils and  to study  the dynamics  of the proof  masses. The  two test
masses are located at the center  of each inertial sensor, and are the
end mirrors of the  OMS.  In fact, one of the test  masses will be the
reference free floating body to perform the translation control of the
spacecraft.  The test masses are made  of an alloy of Pt (27\%) and Au
(73\%), their dimensions are $46\times46\times 46$ mm and their weight
is 1.95  kg.  To  comply with the  top science requirements,  the test
masses must have  certain properties.  For the purpose  of the present
paper  the two most  important ones  are the  magnetic moment  and the
susceptibility,  which  must  be,  respectively,  $|{\bf  m}|<2.0\cdot
10^{-8}$~A~m$^2$ and  $|\chi|<2.5\cdot 10^{-5}$ \cite{bib:trento}. The
volume  of the test  masses is  $V =  0.0046^3$~m$^3$. The  density of
magnetic moment has to be then $|{\bf M}|<9.451 \cdot 10^{-4}$~A/m.

\begin{figure}[t]
 \centering
 \includegraphics[width=0.4\textwidth]{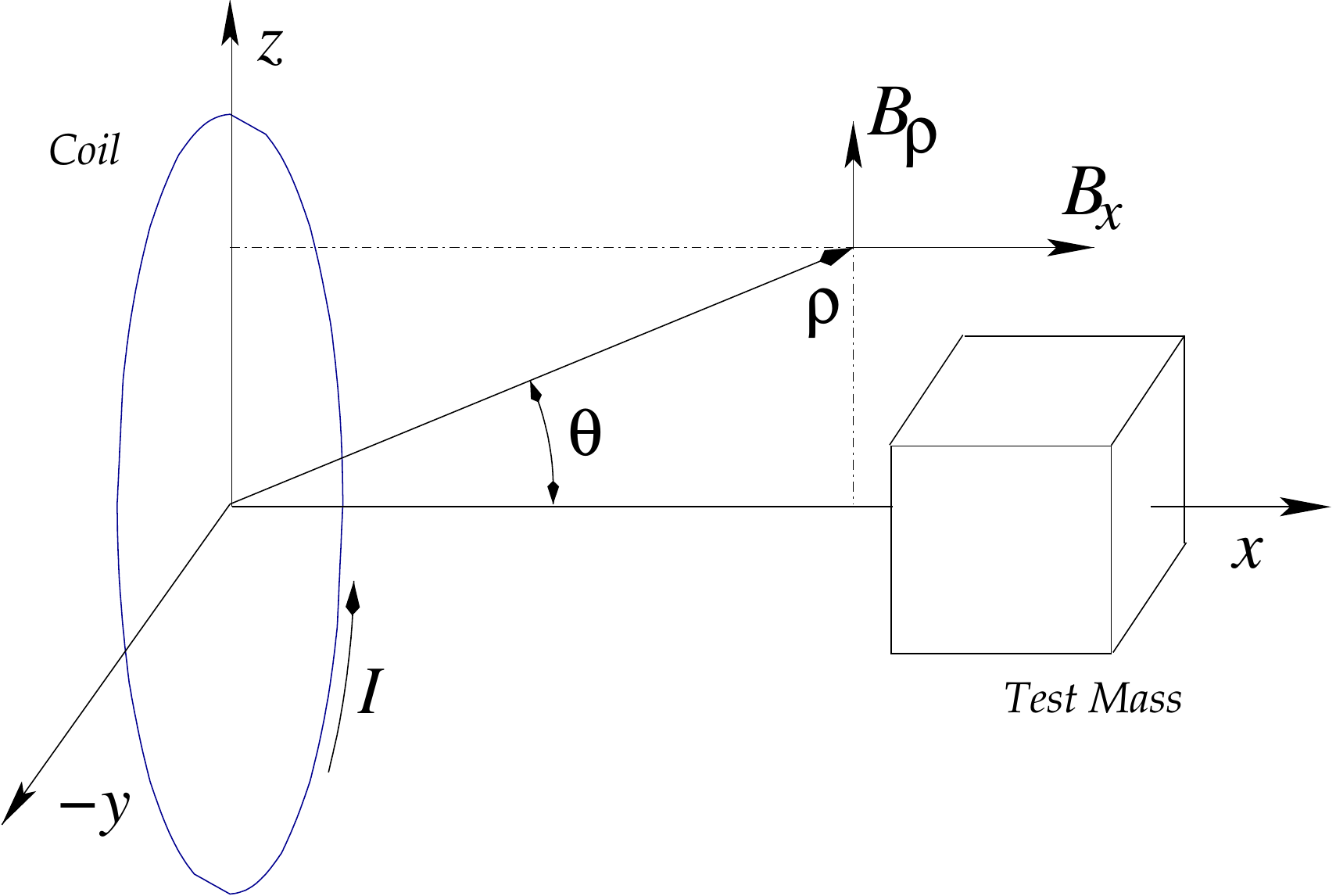}
 \caption{Coordinate reference frame of the experiment.}
 \label{fig.coils}
\end{figure}

The controlled magnetic  field will be produced by  the onboard coils,
which  are   placed  next   to  each  of   the  GRS  towers   ---  see
Fig. \ref{fig.LTP}.   The two circular  induction coils are made  of a
titanium  alloy  (${\rm Ti_6Al_4V}$),  have  $N=$  2\,400 windings  of
radius $r=56.5$  mm. They are placed  85.5 mm away from  the center of
the  respective test  mass.  The  onboard coils  are aligned  with the
$x$-axis  of the  test masses,  thus,  the magnetic  field within  the
volume    of   the    test    masses   has    axial   symmetry,    see
Fig.~\ref{fig.coils}.    If  the   current   fed  to   the  coils   is
$I(t)=I_0\sin\omega  t$,   the  resulting  magnetic   field  (and  its
gradient) will oscillate at the  same frequency.  Thus, when the coils
are switched  on the test masses  rotate and are  displaced from their
equilibrium  positions.  Typical  values  of $I_0$  and $\omega$  are,
respectively, 1~mA and 1~mHz.

\begin{figure}[t]
 \centering
 \includegraphics[width=0.49\textwidth]{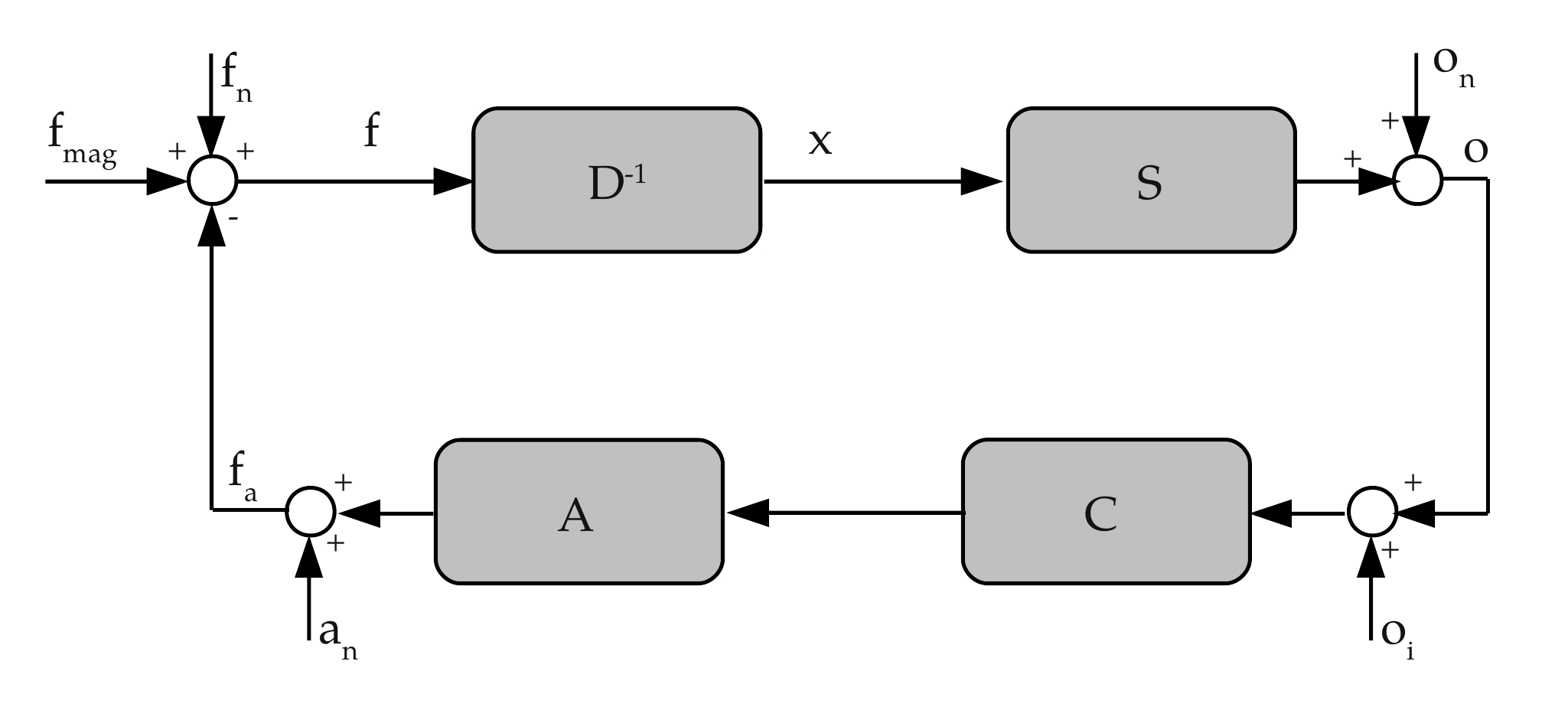}
 \caption{Control system  architecture of LISA  Pathfinder.  ${\bf D}$
          stands for  the dynamical  matrix, ${\bf S}$  represents the
          sensing  matrix  of  the  interferometer, i.e.   the  matrix
          translating the  position of the test mass,  ${\bf x}$, into
          the  interferometer readout, ${\bf  o}$ (${\bf  o_n}$ stands
          for the readout noise) . ${\bf A}$ represents the physics of
          the FEEPS and the electrostatic actuators, and finally ${\bf
          C}$, is  the controller  matrix, implementing the  drag free
          and   low-frequency  control   loops.   ${\bf   o}_{\rm  i}$
          represents the displacement  guidance signals. ${\bf a}_{\rm
          n}$ are  the actuators noise  and ${\bf f}_{\rm a}$  are the
          output forces of the  actuators. ${\bf f}_{\rm mag}$ are the
          magnetic forces  induced by the coils and  ${\bf f}_{\rm n}$
          are the environment force noises disturbing the spacecraft.}
 \label{fig.control}
\end{figure}

Compared to other missions, LTP is a very flexible instrument in terms
of the possible operation scenarios. Nevertheless, we characterize the
magnetic  experiment for  a  fixed operating  mode,  the main  science
mode~\cite{bib:control}. This  mode is a  full 3-dimensional dynamical
mode and  it is schematically shown  in Fig.~\ref{fig.control}.  ${\bf
D}$ is a dynamical matrix which represents the dynamic response of the
spacecraft  and the  test masses  when they  are affected  by specific
forces  (${\bf f}$). This  block consists  of an  18 degree-of-freedom
representation  of the  motion  of these  3  bodies. The  differential
position of the  test masses and their distance  to the spacecraft are
represented by ${\bf x}$ in this block-diagram.

The LTP is endowed with  two different mechanisms to detect the motion
and the  actual position  of the  test masses.  The  first one  is the
interferometer, while the  second is the electrode housing  of each of
the two gravitational reference sensors.   The OMS in LTP is in charge
of  measuring the  distance between  one of  the test  masses  and the
optical  bench,  thus  giving  an  absolute reference,  and  also  the
distance   between  both   test  masses,   providing   a  differential
reading. Due to its ability to perform wavefront sensing, the rotation
angles of the test mass around the $y$- ($\eta$) and $z$-axis ($\phi$)
can  also  be  measured  (the  $y$-  and $z$-axis  in  the  test  mass
coordinate frame are the same axis represented in Fig.~\ref{fig.coils}
but centered at  the test mass). The displacements  are expected to be
measured with  a picometer accuracy  while the rotation angles  can be
measured      with      an       accuracy      of      $\sim      400$
nrad~\cite{bib:interferometer1,bib:interferometer2}.    The  electrode
housing can also be used to determine the position of the test masses.
However, this mechanism only offers readings with nanometer precision.
Consequently,   for  our   application  only   the  readings   of  the
interferometer  will be  used.  The  physical model  of  these sensing
mechanisms  is  included in  ${\bf  S}$,  the  sensing matrix  (figure
\ref{fig.control}). ${\bf o_n}$ is  the readout noise of the different
sensors and ${\bf o}$ the actual measure delivered to the controller.

These  kinematic measurements  are processed  by the  controller block
(${\bf C}$) and  a feedback action is produced on  the LTP dynamics by
the  actuators  (${\bf A}$):  the  micropropulsion  thrusters and  the
electrostatic  actuators. They  produce  an additional  set of  forces
(${\bf f_a}$) with the  following objectives. The drag free controller
acts on the spacecraft  using the micropropulsion thrusters and forces
it to  follow test  mass 1.  The  electrostatic actuators act  on test
mass 2 using the low  frequency suspension, a specific control loop of
very low gain in the LTP measurement bandwidth, that allows to control
in band the differential acceleration  between both test masses at the
same time, avoiding secular drifts or stray motions of the second test
mass.

Finally,  ${\bf f}_{\rm  n}$ represents  the force  noise on  the test
masses, and  ${\bf f}_{\rm mag}$  represents the forces acting  on the
test  masses due  to  the magnetic  field  created by  the coils.  The
complete control architecture can be expressed by the following system
of equations:

\begin{eqnarray}
\label{eq.control}
 {\bf o} & = & {\bf D}^{-1} \cdot {\bf S} \cdot {\bf f} + {\bf o}_{\rm n} 
 \\ \nonumber
 {\bf f} & = & {\bf f}_{\rm mag} + {\bf f}_{\rm n} - {\bf A} \cdot {\bf C} 
 \cdot ({\bf o} + {\bf o}_{\rm i})  -   {\bf a}_{\rm n}\\ \nonumber
\end{eqnarray}

\noindent where  all the  symbols have been  already defined  with the
only exception of ${\bf  o}_{\rm i}$ which represents the displacement
guidance signals of the experiment and ${\bf a}_{\rm n}$ which are the
actuators noise.  Using  Eqs.~(\ref{eq.control}), we can calculate the
transfer function  from the magnetic  forces (${\bf f}_{\rm  mag}$) to
the interferometer readings (${\bf o}$), which turns out to be:

\begin{equation}
 \frac{{\bf o}}{{\bf f_{mag}}} = \frac{{\bf D}^{-1} \cdot 
 {\bf S}}{1 + {\bf D}^{-1} \cdot {\bf S} \cdot {\bf A} \cdot {\bf C}}
\end{equation}

This transfer  function characterizes  the projection of  the magnetic
forces/torques  into  kinematic  motion   of  the  test  masses.   The
controllers have  been designed to deliver very  sensitive readings of
the  differential  motion  of  both  test masses  between  1\,mHz  and
30\,mHz, the measurement bandwidth of the LTP mission. For simplicity,
if we consider only the one-dimensional model:

\begin{eqnarray}
 {\bf D}  & = & \left(
\begin{array}{cc}
s^2 + \omega^2_1 & 0 \\
\omega^2_2 - \omega^2_1 &  s^2 + \omega^2_2 \\
\end{array}
\right) ,
\\ \nonumber
 {\bf S}  & = & \left(
\begin{array}{cc}
1 & 0 \\
\delta_{12} & 1 \\
\end{array}
\right) , 
\\ \nonumber
 {\bf C}  & = & \left(
\begin{array}{cc}
C_{\rm DF} & 0 \\
0 & C_{\rm LFS} \\
\end{array}
\right) , 
\\ \nonumber
{\bf A}  & = & \left(
\begin{array}{cc}
A_{\rm FEEP} & 0 \\
0 & A_{\rm EA} \\
\end{array}
\right)
\\ \nonumber
\end{eqnarray}

\noindent where $\omega_1$ and $\omega_2$ are the stiffness parameters
coupling the motion of each test mass to the motion of the spacecraft,
$\delta_{12}$  is the  interferometer  channel crosscoupling.  $C_{\rm
DF}$  and  $C_{\rm LFS}$  are  the drag  free  and  the low  frequency
suspension  controller transfer  functions respectively  and, finally,
$A_{\rm FEEP}$ and  $A_{\rm EA}$ are the physical  models for the FEEP
thrusters and the electrostatic actuators.

\section{Forces and torques}
\label{sec:forces}

Using Eq.~(\ref{eq.1})  and neglecting the  environmental field (which
is much smaller than the  applied field, 500 nT), the $x$-component of
the force acting on the test mass is

\begin{equation}
 F_x  =  \left\langle{\bf M}\Cdot\Nabla B_{0,x}\right\rangle 
 V\,\sin\omega t 
  +  \frac{\chi V}{\mu_0}\,\left\langle
 {\bf B}_0\Cdot\Nabla B_{0,x}\right\rangle\,\sin^2\omega t
\label{eq.forces}
\end{equation}

\begin{figure}[t]
 \centering
 \includegraphics[width=0.4\textwidth]{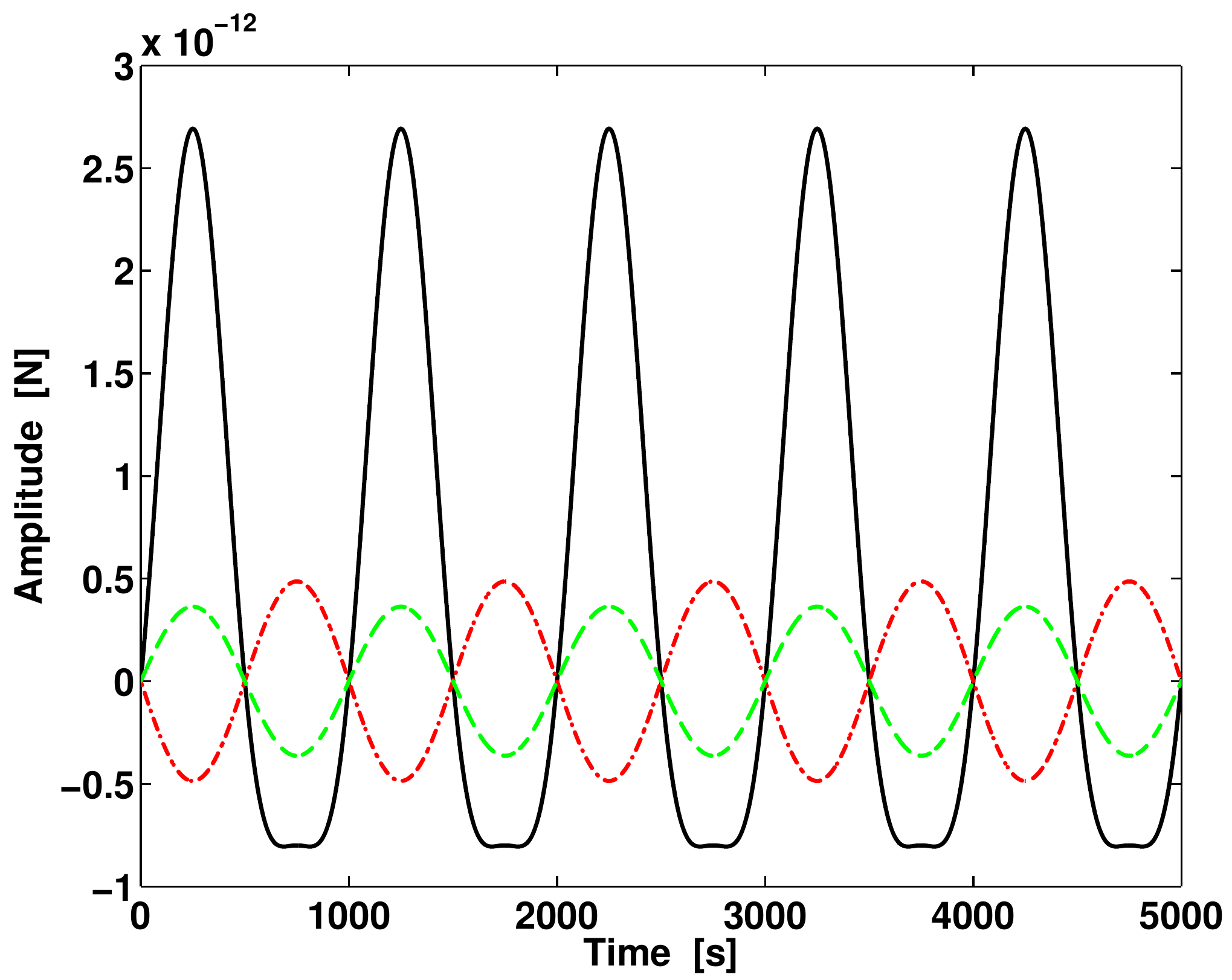}
 \caption{The three components of the force on test mass 1 when coil 1
          is  on. These  forces depend  on  the values  of $\chi$  and
          $M_x$, $M_y$  and $M_z$. For  this specific example  we have
          adopted $M_x  = 16.4 \cdot  10^{-5}\,{\rm A/m}$, $M_y  = 9.1
          \cdot 10^{-5}\,{\rm  A/m}$, $M_z =  -6.8 \cdot 10^{-5}\,{\rm
          A/m}$  (a random  orientation of  the maximum  {\bf  M}) and
          $\chi = 2.5 \cdot 10^{-5}$.   $I_0$ is 1\,mA and $\omega$ is
          1\,mHz. The $x$-component  of the force is shown  as a solid
          black  line, whereas  the  $y$-component is  displayed as  a
          dashed-dotted red line and the $z$-component is displayed as
          dashed green line.}
 \label{fig.force}
\end{figure}

\noindent where {\bf B}$_0$ is  the field produced by the coils. Thus,
since $\sin^2\omega  t=(1-\cos 2\omega t)/2$,  the linear acceleration
of the test  masses has two separate frequencies,  one at $\omega$ and
the other  one at $2\omega$,  and also a  DC component.  The  force on
test mass 1  is plotted in figure \ref{fig.force}.   The torque acting
on the  test mass also  has a similar  behavior.  However, it  must be
noted that, because of the symmetry of the applied magnetic field, the
torque only has one frequency component:

\begin{equation}
 \mathbf{N} = \left\langle\mathbf{M}\times\mathbf{B}_0 +
 \mathbf{r}\times\left[\left(\mathbf{M}\Cdot\Nabla\right)\mathbf{B}_0\right] 
 \right\rangle V\,\sin\omega t
 \label{eq.torques}
\end{equation}

\begin{figure}[t]
 \centering
 \includegraphics[width=0.4\textwidth]{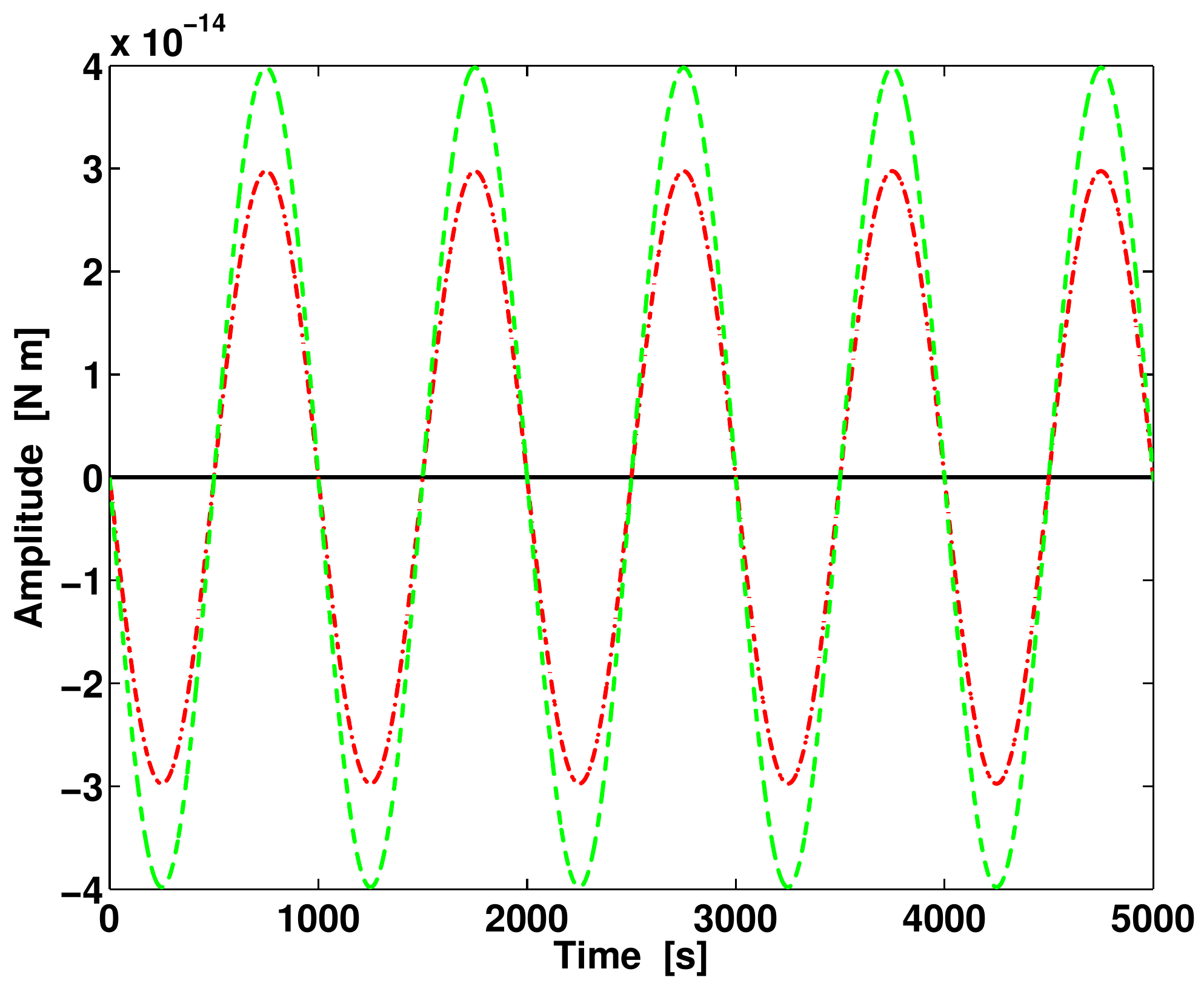}
 \caption{Torques  on test  mass 1  when coil  1 is  on.  The magnetic
          properties of the  test mass are the same  adopted in figure
          \ref{fig.force}.  The  $x$-, $y$- and  $z$-components of the
          torque are  shown using  black solid, red  dashed-dotted and
          green dashed lines, respectively.}
 \label{fig.torque}
\end{figure}

The resulting torques are displayed in figure \ref{fig.torque}.  It is
important  to realize  that only  the $y$-  and $z$-components  of the
torque can  be measured with  the interferometer, as $N_x$  produces a
rotation  around   the  direction   of  the  laser   beam.   Moreover,
decomposing Eqs.  (\ref{eq.forces}) and (\ref{eq.torques}), it is easy
to show  that the $x$-component of  the force on the  test mass (which
can be obtained from its displacement) and the $y$- and $z$-components
of the torque (which can be  obtained from the rotation angles) can be
cast in the form:

\begin{eqnarray}
\label{eq.structure}
 F_x &  = &\chi \cdot f_{x_{\rm DC}} + M_x \cdot f_{x_{1\omega}} + 
           \chi \cdot f_{x_{2\omega}} \nonumber \\
 N_y &  = & M_z \cdot f_{y_{1\omega}} \\ 
 N_z &  = & M_y \cdot f_{z_{1\omega}} \nonumber 
\end{eqnarray}

\noindent   where   $f_{x_{\rm   DC}}$   is   a   constant   function,
$f_{x_{1\omega}}$,  $f_{y_{1\omega}}$ and  $f_{z_{1\omega}}$ oscillate
at  $\omega$  and $f_{x_{2\omega}}$  oscillates  at $2\omega$.  Hence,
$N_y$  and   $N_z$  will  be   used  to  estimate  $M_z$   and  $M_y$,
respectively, while  the differential displacement of  the test masses
will be used to measure $M_x$ and~$\chi$.

\section{Modelling of the noise sources}
\label{sec:NoiseSources}

The  forces   and  torques   shown  in  figures   \ref{fig.force}  and
\ref{fig.torque} correspond to an ideal case. However, the real forces
and   torques    acting   on   the    test   masses   will    not   be
noise-free. Additionally,  the outputs detected  by the interferometer
will also be  affected by several noise sources.   Thus, to assess the
feasibility of the experiment we need to model the noise sources. This
is precisely the aim of this section.

\subsection{Magnetic hardware noise}

The  stability of  the magnetic  field ($S^{1/2}_B$)  produced  by the
coils  at   the  position  of   the  test  masses  and   its  gradient
($S^{1/2}_{\partial{B}_x/\partial  x}$) must be,  respectively, better
than       5~nT~Hz$^{-1/2}$       and      12~nT~m$^{-1}$~Hz~$^{-1/2}$
\cite{bib:trento}  within the  measurement  bandwidth (1~mHz  $< f  <$
30~mHz).  This can  be translated into a requirement  on the stability
of the  injected current.   It turns out  that the requirement  on the
magnetic field gradient is the  more demanding one, and using Ampere's
law it is  straightforward to show that it is  equivalent to a current
fluctuation ($S^{1/2}_I$) requirement of 110~nA~Hz$^{-1/2}$ within the
measurement bandwidth.

\begin{figure}[h]
 \centering
 \includegraphics[width=0.45\textwidth]{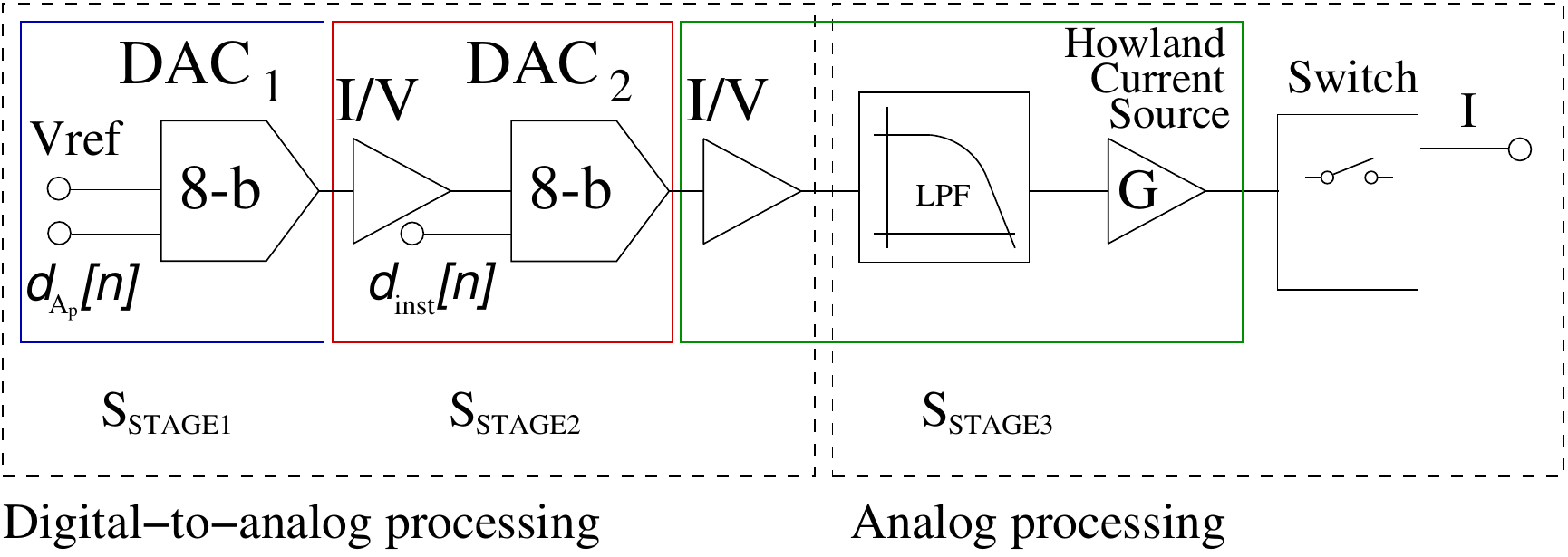}
 \caption{Block diagram of the coil's current source.}
 \label{fig.curr.source}
\end{figure}

In  figure~\ref{fig.curr.source},  we  show  a block  diagram  of  the
different hardware stages  used to produce the current  that feeds the
coils.   This  block  diagram   has  two  parts:  a  digital-to-analog
processing  stage and  an analog  processing stage.   The  first block
contains   two  digital-to-analog   converters   (DAC),  followed   by
transresistance  amplifiers  (I/V)  delivering  low  impedance  output
voltages. The first  DAC sets the reference voltage  of the second DAC
and thus the  peak amplitude of the sinusoidal  current applied to the
coil. The second one is  configured with bipolar operation to generate
the quantized  signal with the previously selected  amplitude.  In the
analog processing block the  signal is low-pass-filtered and amplified
with a Howland current  source~\cite{bib:franco}. Finally, a switch is
used  to  select one  of  the  three  possible states:  short-circuit,
open-circuit or connected. The noise of the current source chain for a
DC signal, $S_{I_{\rm DC}}^{1/2}$, can be written as:

\begin{equation}
S_{I_{\rm DC}}^{1/2} (I,\omega) \simeq 
\left[G_2  G_3  S_{{\rm STAGE}_1} + 
      G_3 S_{{\rm STAGE}_2} +
      S_{{\rm STAGE}_3}\right]^{1/2}
\end{equation}

\noindent  where  $S_{{\rm STAGE}_1}$  is  the  noise  density of  the
voltage reference and the first  DAC, $S_{{\rm STAGE}_2}$ is the noise
density of  the transimpedance amplifier and the  second DAC, $S_{{\rm
STAGE}_3}$  is  the noise  density  of  the transimpedance  amplifier,
low-pass filter and  Howland current source, and $G_2$  and $G_3$ are,
respectively, the gains in the the second and third stages.  The noise
density of  the first stage can  be obtained as a  contribution of the
first DAC, $S_{{\rm DAC}_1}$, and of the reference voltage, $S_{V_{\rm
ref}}$.   However,  the  noise   contribution  of  the  first  DAC  is
negligible  with respect  to  that of  the  reference source,  $V_{\rm
ref}$.  Thus, if  the DAC works at its  full scale, $S_{{\rm STAGE}_1}
\sim S_{V_{\rm ref}}$.  Likewise,  $S_{{\rm STAGE}_2}$ can be computed
as  a  combination of  two  different  sources  $S_{{\rm DAC}_2}$  and
$S_{\rm I/V}$, being $S_{{\rm DAC}_2}$ the noise density of the second
DAC  and $S_{\rm  I/V}$ that  of the  transresistance  amplifiers, and
finally  $S_{{\rm STAGE}_3}=S_{\rm  LPF}+S_{\rm  HCS}$, where  $S_{\rm
LPF}$ and $S_{\rm HCS}$ are,  respectively, the noise densities of the
low-pass  filter  and of  the  Howland  current  source.  These  noise
sources  have been  measured for  a DC  current of  1~mA.   The values
obtained  are $S_{\rm  STAGE_1}=9.6\, {\rm  nA  \,Hz}^{-1/2}$, $S_{\rm
STAGE_2}=0.03\, {\rm nA \,Hz}^{-1/2}$, and $S_{{\rm STAGE}_3}= 26.4 \,
{\rm  nA  \,Hz}^{-1/2}$,  respectively.   The  details  of  how  these
measurements were done are out of  the scope of this paper and will be
provided elsewhere.   However, we  mention that although  the previous
analysis  has been  performed  for  DC currents,  when  we operate  at
1\,~mHz the  dominant noise  source is the  quantization noise  of the
second DAC. This means that the total current noise can be modeled as:

\begin{equation}
S_{I_{\rm AC}}^{1/2} \simeq \left[G_3 S_{{\rm DAC}_2}\right]^{1/2}
\end{equation} 

\noindent where the sole contributor  is the quantization noise of the
second  stage,   and  thus  the   total  noise  assuming   an  uniform
quantization and  a signal amplitude greater than  a quantization step
is given by:

\begin{equation}
 S_{I_{\rm AC}}^{1/2}\simeq\frac{2I_0}{2^{N_{\rm b}}}\frac{1}
    {\sqrt{12\cdot f_{\rm s}}}
\label{quant}
\end{equation} 

\noindent where  $f_{\rm s}$ is the sampling  frequency, $N_{\rm b}=8$
is the  number of bits  of the onboard  hardware, and the rest  of the
symbols have been already defined.

\begin{figure}[t]
 \centering
 \includegraphics[width=0.4\textwidth]{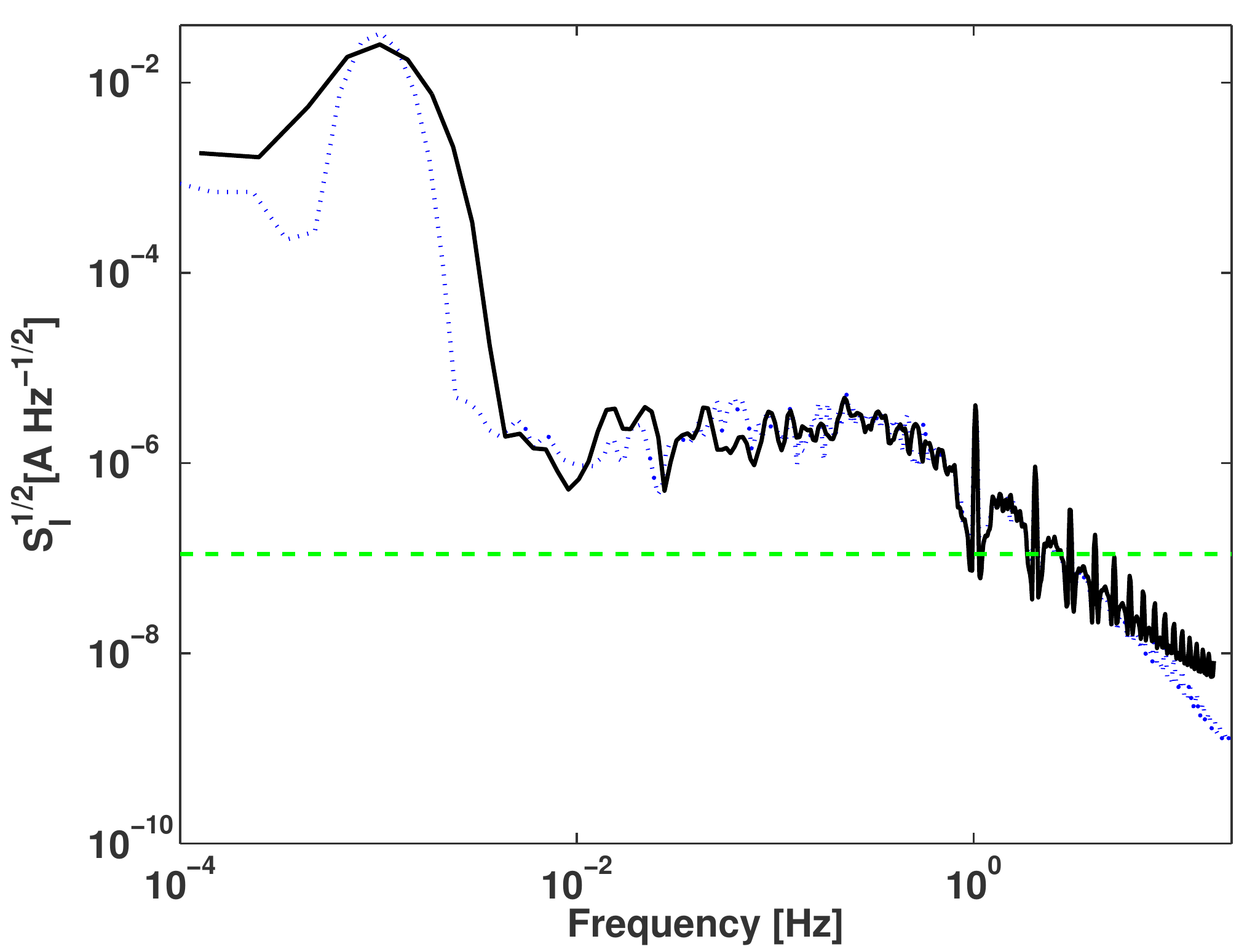}
 \caption{Noise  density  for a  1~${\rm  mA}$  sinusoidal current  of
          1~${\rm mHz}$ nominal frequency.  The dotted blue line shows
          the measured noise density,  whereas the solid black line is
          our theoretical estimate (see  text for details).  The green
          dashed   line  represents   the   DC  current   fluctuations
          requirement.}
 \label{fig.ACcurrentStability}
\end{figure}

The  DMU delivers  the output  at a  rate of  1024 samples  per cycle.
Thus,  the 1~mHz  sinusoidal signal  will  be sampled  at a  frequency
$f_{\rm s}=1.024~$Hz.   Since the  highest sinusoidal current  used in
the experiment will be 1\,mA, the highest noise will be 2.22~$\mu {\rm
A}\,{\rm Hz}^{-1/2}$.  This noise is 2 orders of magnitude larger than
all  other contributions  and therefore  will be  the  dominant noise.
Figure \ref{fig.ACcurrentStability} shows the current noise for coil 1
for a signal of 1\,mHz and  1\,mA (which are the nominal values of the
experiment). As can be seen, the quantization noise level is above the
DC current  stability requirement for the mission  (dashed green line)
but, as  it will  be shown  below, it is  still sufficiently  small to
allow a  reliable estimation  of the magnetic  properties of  the test
masses.  It  is also interesting  to note that the  quantization noise
(thus the  total noise) can be reduced  when a 16-bit DAC  is used ---
see  Eq. (\ref{quant}).  Finally,  note that  our theoretical  results
nicely match the experimental results.

\subsection{Other noise sources}

So  far   we  have  discussed   the  noise  density  of   the  onboard
coils. However,  the different electronic subsystems  of the satellite
produce  magnetic fields, which  are also  noisy.  The  magnetic noise
produced by these subsystems  ($\sim 50$) has been modeled considering
the fluctuating  values of their  magnetic moments~\cite{bib:myPaper}.
The expected  displacement noise spectral density at  the positions of
the  test  masses due  to  the  magnetic  field generated  by  several
electronic   boxes   is   3   $\cdot$  10$^{-11}$   m~Hz$^{-1/2}$   at
1\,mHz.

Other noise sources that affect the measurements are the readout noise
of the OMS, the noise induced by the GRS and that of the star tracker.
As   already   mentioned,   we   only   use  the   readings   of   the
interferometer. However,  the control chain  uses the readings  of the
inertial sensors  and of the star  tracker to stabilize  the drag free
motion and  the satellite attitude.   Therefore, both the GRS  and the
star tracker affect  the readings of the OMS,  and must be considered.
The  dominant  noise is  the  readout  noise  in displacement  of  the
interferometer.  We  model it  using a two  pole/two zero  noise shape
filter~\cite{bib:dfacsICD}:

\begin{equation}
\left|N_{\rm OMS}\right|=\left(\frac{f+2\cdot10^{-2}/(2\pi)}
			  {f+2\cdot10^{-4}/(2\pi)}\right)^{2}
\end{equation}

\noindent At  the positions of the  test masses this results  in a low
frequency     displacement    noise     of     around    5     $\cdot$
10$^{-11}$~m~Hz$^{-1/2}$ at  1\,mHz but it is  the largest contributor
at high frequency.

\begin{figure}[t]
\centering
    \subfigure{\includegraphics[width =0.4\textwidth]{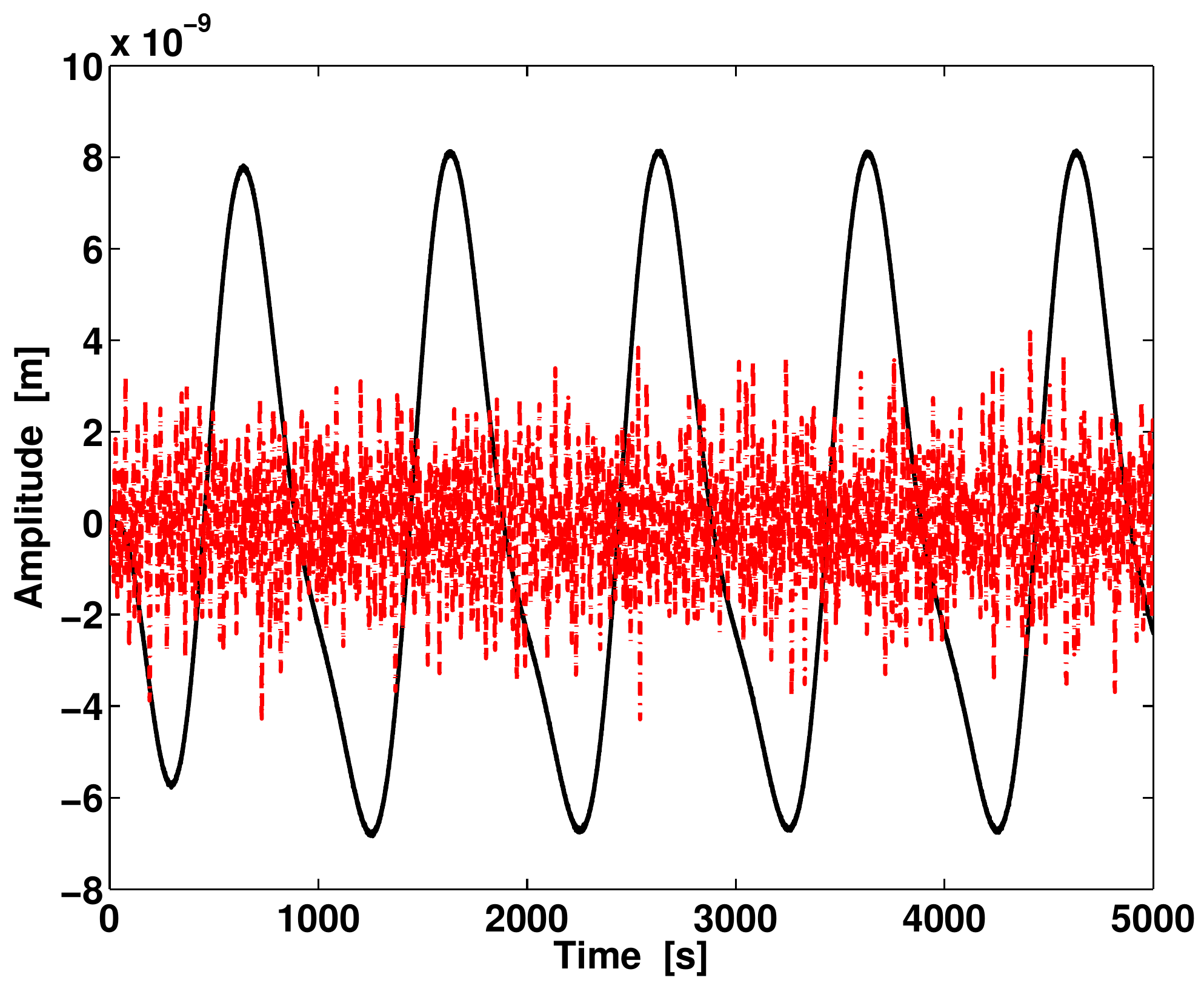}}
    \subfigure{\includegraphics[width =0.4\textwidth]{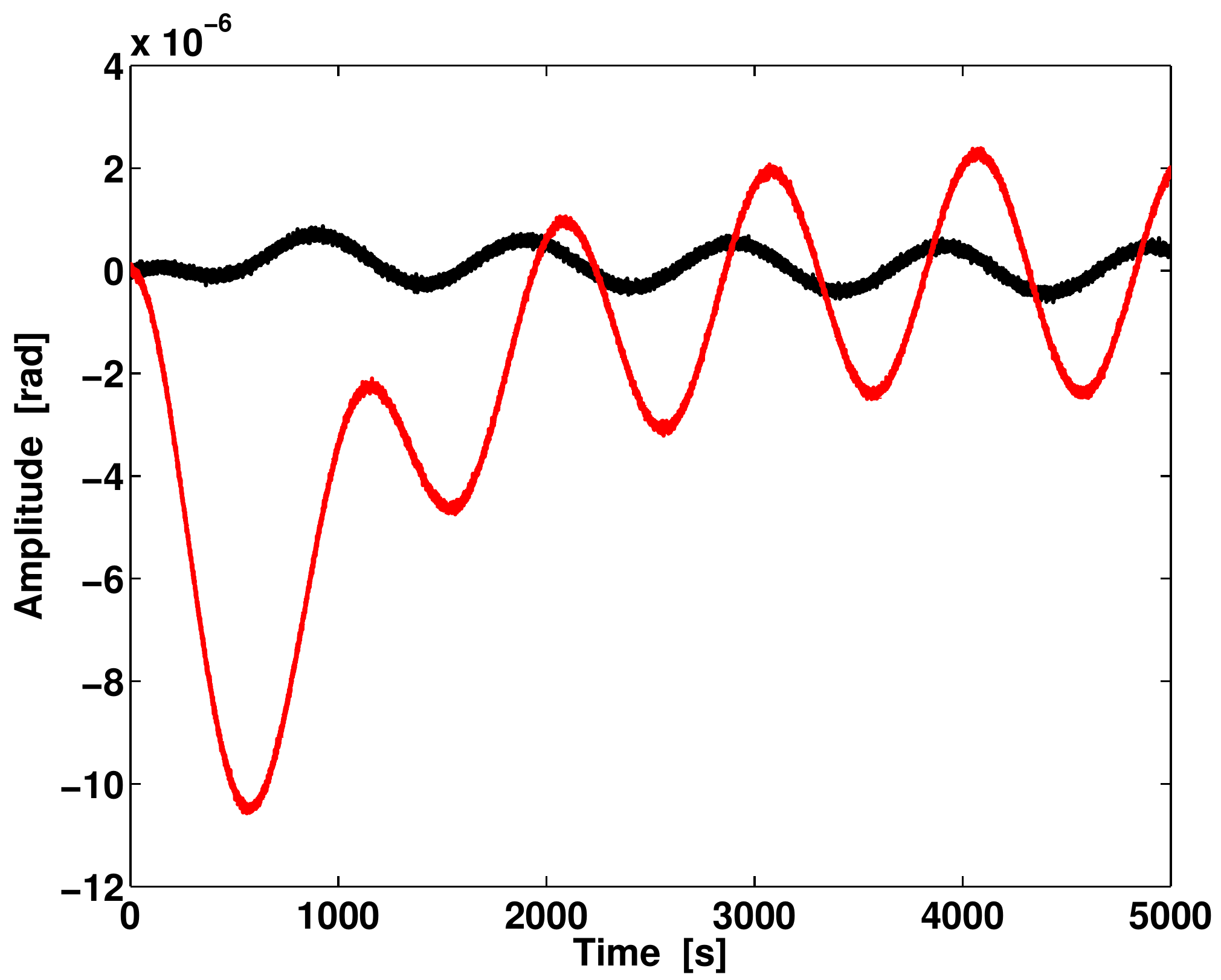}}
    \caption{Top  panel:  relative  displacement  of the  test  masses
             measured   using   the   differential  channel   of   the
             interferometer (solid  black line).  The  dashed red line
             shows  the  absolute  displacement  with respect  to  the
             optical bench.  Bottom panel: rotation about the $y$-axis
             --- black  solid  line  ---  and  the  $z$-axis  ---  red
             line. Both panels  show the response of test  mass 1 when
             only coil 1 is fed with a current of 1\,mA and 1\,mHz.}
\label{fig.displacement}
\end{figure}

The   actuators   are   also   a   relevant  noise   source   at   low
frequencies. This  noise is  due to the  capacitive actuators  and the
FEEP thrusters which  have to ensure the free fall  motion of the test
mass. The micro-propulsion system of LISA Pathfinder is composed of 12
FEEP thrusters. The force noise of each individual thruster is modeled
as~\cite{bib:dfacsICD}:

\begin{equation}
\left|N_{\rm FEEP}\right|=\left(\frac{f+ 10^{-2}}
                          {f+ 10^{-3}}\right)^{2}
\end{equation}

\noindent Their  contribution in  the differential readout  channel is
around  2  $\cdot$  10$^{-10}$  m~Hz$^{-1/2}$  at 1\,mHz,  and  it  is
somewhat larger  on the absolute channel (which  provides the distance
from test mass  1 to the spacecraft) where the  FEEP thrusters are the
main disturbance source~\cite{bib:trento}.

Solar  and  infrared  disturbances  have  been  also  modeled.   Solar
disturbances are  due to the solar  flux impacting on  the surfaces of
the spacecraft. Infrared disturbances are due to the infrared emission
from the  spacecraft external surfaces.   These two noise  sources are
the  most  important contributors  of  disturbance  on the  spacecraft
coordinates but they are highly attenuated by the control architecture
and they  turn out to  be completely negligible in  the interferometer
readings.   Finally, the  test  mass disturbance  noise represent  the
internal disturbances at the  test mass coordinates. They are expected
to be the most relevant sources at low frequency on the interferometer
readings and their contribution has been estimated to be $\sim 3 \cdot
10^{-10}$~m~Hz$^{-1/2}$ at  1~mHz.  All these sources  of noise affect
the quality  of the  estimate of the  magnetic properties of  the test
masses, as it will be shown below.

\section{Results}
\label{sec:Results}

We have estimated the magnetic properties of the test masses using the
readings   provided   by  the   mission   telemetry.   The   telemetry
corresponding to the magnetic experiments will consist of the commands
sent  to the coils,  the displacement  readings of  the interferometer
(namely,  the absolute  and differential  readings) and  the wavefront
rotation readings about the $y$- ($\eta$) and $z$- axes ($\phi$).  For
simplicity, we have assumed that the stiffness of the test masses, the
actuators  gains  and the  interferometer  crosscoupling factors  have
already been determined~\cite{bib:MDC}.

The  simulated displacements  and  rotations measured  by the  onboard
interferometer are displayed  in figure \ref{fig.displacement}.  These
displacements  and  rotations   have  been  obtained  integrating  the
equations of motion of a rigid  solid, and including the drag free and
low frequency  controllers --- see  Sect.  \ref{sec:Experiment}.  Note
that the problem has 18 degrees of freedom. In particular, each of the
two test masses has 6 degrees  of freedom, and the spacecraft also has
6 degrees  of freedom.  The  closed loop simulation is  performed with
appropriate   simulation  tools   that  will   be  used   for  mission
operations~\cite{bib:DataAnalysis,  bib:LTPDA}. As  can  be seen,  the
displacements of  the test masses are  below 8 nm,  while in permanent
regime  the  corresponding  rotations  have amplitudes  of  $\sim  4\,
\mu$rad.  The very  long transient of about 3\,000\,s  of the rotation
excusions ---  see the  bottom panel of  figure \ref{fig.displacement}
--- is  due  to the  effect  of  the  low-frequency controller.   This
controller is  designed to avoid  drift excursions of the  test masses
with frequencies smaller than  1\,mHz.  Consequently, the transient is
very long.  A similar transient,  although less evident, is present in
the   differential  reading   ---  see   the  top   panel   of  figure
\ref{fig.displacement}.  The  reading of the  displacement channel has
two  frequency  components,  $\omega$  and $2\omega$,  however,  these
components  are  difficult  to  see   in  the  time  series  shown  in
figure~\ref{fig.displacement}.  Additionally, these two components are
not in  phase with the forces shown  in figure~\ref{fig.force} because
the LPF dynamics  and the controllers introduce a  phase delay to each
of  the two  components. Finally,  it is  worth mentioning  that these
displacements and rotations are within acceptable margins because they
do not exceed the authority  limits of the drag free and low-frequency
controllers.

\begin{table*}[t!]
\caption{Errors in the estimates of the magnetic properties.}
\label{tab:estimation}
\begin{center}
\begin{tabular}{lcccc}
\hline
\hline
          & $\Delta\hat{M}_x$ & 
            $\Delta\hat{M}_y$ & 
            $\Delta\hat{M}_z$ & 
            $\Delta\hat{\chi}$ \\ 
\hline \\

 No noise                     & $10^{-13}\%$ & $10^{-13}\%$ & $10^{-13}\%$ & $10^{-13}\%$ \\
 Hardware noise               & 0.13\%      & 0.08\%     & 0.09\%      & 0.12\%      \\
 Environmental noise          & 0.12\%      & 0.26\%     & 0.24\%      & 0.10\%      \\
 Sensors noise                & 0.87\%      & 0.97\%     & 1.05\%      & 1.01\%      \\
 Actuators noise              & 0.96\%      & 0.99\%     & 1.25\%      & 1.17\%      \\
 Solar and infrared noise     & 0.03\%      & 0.02\%     & 0.05\%      & 0.06\%      \\
 Test mass disturbance noise  & 0.82\%      & 0.73\%     & 0.75\%      & 0.99\%      \\ 
 All sources                  & 1.15\%      & 1.53\%     & 1.72\%      & 1.25\%      \\ 
\hline
\hline
\end{tabular}
\end{center}
\end{table*}

\begin{figure}[t!]
 \centering
 \includegraphics[width=0.4\textwidth]{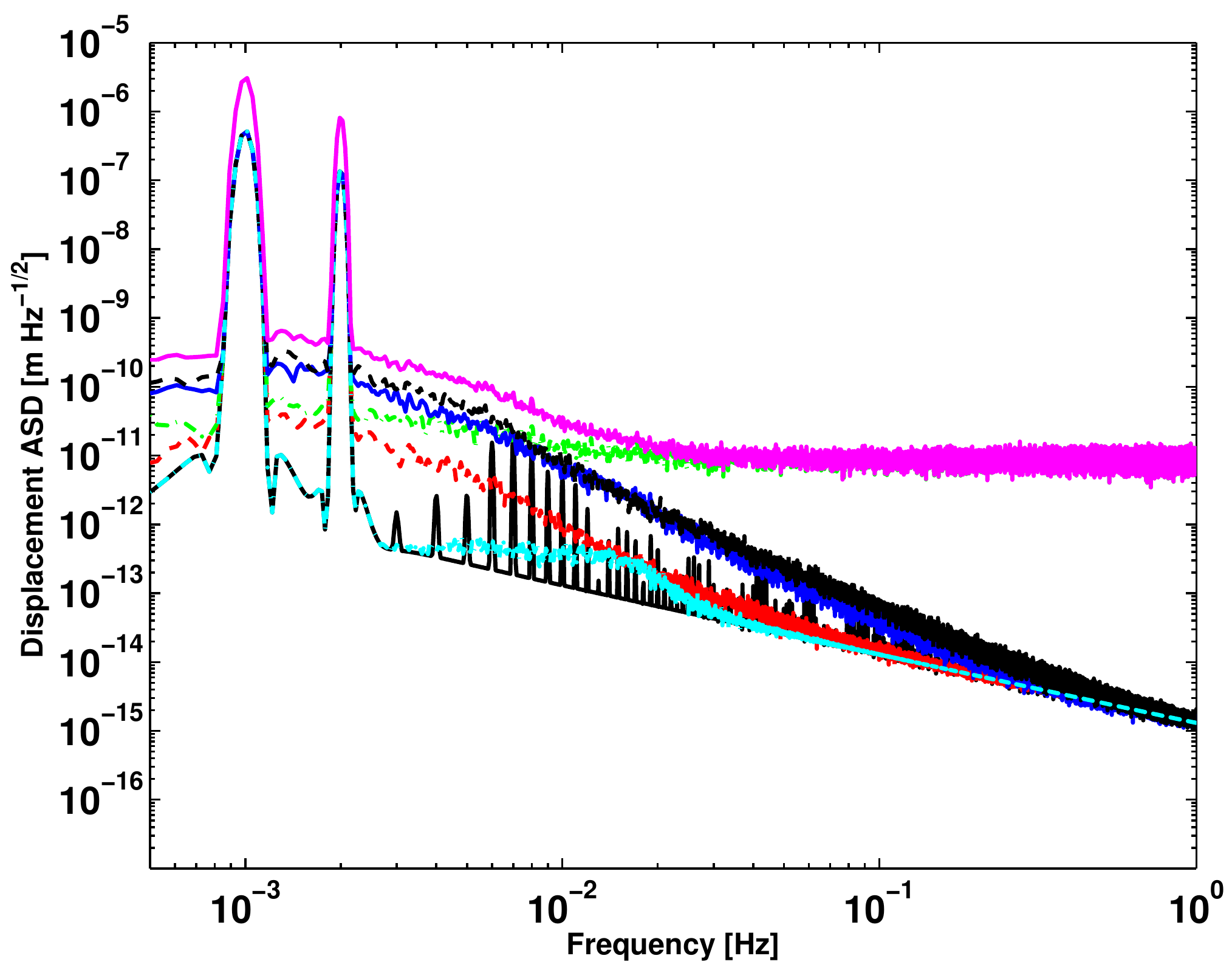}
 \caption{Noise  breakdown  of  each  of  the  noise  sources  to  the
          differential displacement  signal received when  a 1\,mA and
          1\,mHz current  is circulating in  coil 1.  The  black solid
          line  shows  the result  of  considering  only the  magnetic
          hardware  noise.  The red  dashed line  shows the  result of
          considering only the  environmental magnetic noise. The cyan
          dashed-dotted  line only  takes into  account the  solar and
          infrared emission noise. The green dotted green, the sensors
          contribution.  The  solid  blue  line the  actuators  noise,
          whereas  the   black  dashed   line  shows  the   result  of
          considering the test mass  noise.  Finally, in magenta solid
          line the result of considering all the noise sources.}
 \label{fig.noiseBreakdown}
\end{figure}

To  further   illustrate  the   feasibility  of  the   experiment,  in
figure~\ref{fig.noiseBreakdown}  we show  the noise  breakdown  of the
differential displacement reading  of the interferometer.  This figure
has been obtained  simulating the output of the  entire instrument for
each   of   the   noise   sources  presented   previously   in   Sect.
\ref{sec:NoiseSources}. We simulated  100\,000 seconds for each source
of noise. Then,  we performed the spectral estimation  with a smoothed
power  spectral  estimator  based  on  the  Welch  estimator  using  a
Blackman-Harris  window~\cite{bib:welch}. The  time  domain simulation
and the  spectral analysis have  been performed using the  {\tt LTPDA}
toolbox \cite{bib:LTPDA}. This is the  data analysis tool that will be
used for mission operations.  As  can be seen, in the frequency domain
the signals at $\omega$ and  $2\omega$ are clearly visible.  Note that
the most important contribution is that of the sensors noise, which is
mainly   characterized   by    the   high-frequency   noise   of   the
interferometer,  the  contribution  of  the  FEEP  thrusters  and  the
disturbance  noise of the  test masses.   The environment  noise, the
magnetic  hardware  noise  and  of  the solar  and  infrared  emission
contribution are totally negligible in the interferometer readings and
do  not represent  any restriction  in terms  of  parameter estimation
quality.

We have already shown that the displacements and rotations of the test
masses can be detected even in the case in which all the noise sources
are considered. Now  the question to be answered  is to which accuracy
the magnetic properties  of the test masses can  be estimated. To this
end    we   have    used    a   classical    linear   least    squares
procedure~\cite{bib:leastSquares}.    The   magnetic  parameters   are
estimated  in  the  following  way.   Let $D_x$  be  the  differential
displacement signals from the  interferometer, and $R_y$ and $R_z$ the
rotation    excursions   around   the    $y$-   and    the   $z$-axis,
respectively, we write then:

\begin{eqnarray}
D_x & = & 
\left(\begin{array}{cc}
 d_{{x_{1\omega}}} & d_{{x_{2\omega}}}
\end{array}\right) \cdot
\left(\begin{array}{c}
 M_x \\ \chi
\end{array}\right) + n_{d_x} \nonumber \\ 
R_y &=& M_z \cdot r_{{y_{1\omega}}} + n_{r_y} \\
R_z &=& M_y \cdot r_{{z_{1\omega}}} + n_{r_z}\nonumber \\ \nonumber
\end{eqnarray}

\noindent where we have used Eq.~(\ref{eq.structure}), $d$ and $r$ are
the  signals in  displacement  and rotation  matched  to the  expected
waveforms    in   $\omega$   and    $2\omega$,   as    obtained   from
Eq.~(\ref{eq.structure}), and  $n_{d_x}$, $n_{r_y}$ and  $n_{r_z}$ are
the errors  of the estimation  model, namely, the  displacement error,
the $\eta$-  and the $\phi$- error, respectively.   Then the estimated
magnetic properties of the test  masses ($\hat M_x$, $\hat M_y$, $\hat
M_z$ and  $\hat \chi$) applying  least square techniques  are computed
as:

\begin{eqnarray}
\label{eq:estimation1}
\left(
\begin{array}{c}
  \hat{M_x} \\
  \hat{\chi}
\end{array}
\right) &
= &
\left[
\left(\begin{array}{c}
 {d_{{x_{1\omega}}}}^{\rm T} \\ {d_{{x_{2\omega}}}}^{\rm T}
\end{array}\right) \cdot (\begin{array}{cc}
 d_{{x_{1\omega}}} & d_{{x_{2\omega}}}
\end{array})
\right]^{-1}
\left( 
\begin{array}{c}
d_{{x_{1\omega}}} \\
d_{{x_{2\omega}}}
\end{array}\right) D_{x}
\nonumber \\
\hat{M_z} & = & \left[
{r_{{y_{1\omega}}}}^{\rm T} \cdot r_{{y_{1\omega}}} 
\right]^{-1} {r_{{y_{1\omega}}}}^{\rm T} \cdot  R_{y}
\\
\hat{M_y} & = & \left[
{r_{{z_{1\omega}}}}^{\rm T} \cdot r_{{z_{1\omega}}} 
\right]^{-1} {r_{{z_{1\omega}}}}^{\rm T} \cdot  R_{z}
\nonumber 
\end{eqnarray}

As can  be seen, the values  of $M_x$ and $\chi$  must be disentangled
from  $D_x$  because  the  dynamics   of  the  test  masses  show  two
frequencies, while  $R_y$ and $R_z$  can be directly used  to estimate
the values of $M_z$ and $M_y$.

We have examined the contribution of  each of the noise sources in our
estimation accuracy. The accuracy  of the measurements of the magnetic
properties  of the  test masses  is  mainly affected  by the  specific
contribution of the noise  source within the measurement bandwith. For
instance,  if some  noise source  has a  relevant  contribution around
1\,mHz  in the  rotation  signals, the  estimating  algorithm can  not
disentangle this  contribution from that  of the injected  torque.  In
table  \ref{tab:estimation} we  list the  accuracies of  the estimated
magnetic parameters obtained for  each of the individual noise sources
and  that  obtained when  all  the  noise  sources are  present  (last
row). Rather naturally, the results for $M_x$ and $\chi$ shown in this
table are closely  related to the noise contributions  shown in figure
\ref{fig.noiseBreakdown}            ---            note           that
figure~\ref{fig.noiseBreakdown},  only shows  the noise  breakdown for
the differential displacement channel. The largest contribution to the
error budget are, as expected, the actuators noise, the interferometer
noise  and  the test  mass  disturbances.   Nevertheless, the  overall
quality of the  estimate is fairly good, 1.43\%  (mean square error of
the  relative   errors  of  all  the  estimated   parameters).  It  is
interesting to note as well that even if the rotations signals present
signal to noise ratios around a  factor of 3 smaller, we obtain errors
of  the  same  order of  magnitude  for  the  estimates of  $M_y$  and
$M_z$. This stems  from the fact that the signals  from which they are
obtained must not be disentangled.

\section{Conclusions}
\label{sec:Conclusions}

In  this paper  we  have  confirmed the  feasibility  of deriving  the
magnetic  properties  of the  test  masses  of  LISA Pathfinder.   The
magnetic  experiment  is  based  on  injecting  controlled  sinusoidal
currents through the  on-board coils and studying the  dynamics of the
test masses, as measured with  the optical metrology subsystem. In our
study we have performed numerical calculations that incorporate a full
model of the dynamics of  the test masses, realistic noise sources and
up-to-date simulations of the interferometer and inertial sensors.  In
particular, all  the degrees of freedom  of the test  masses have been
appropriately  analyzed  and we  have  fully  taken  into account  the
control architecture  of LISA Pathfinder.   We have obtained  that the
displacements of the test masses  along the $x$-direction are $\sim 8$
nm,  while the rotation  excursions are  approximately $4  \, \mu$rad.
These  findings confirm  that the  magnetic experiment  is  within the
authority  margins  of the  drag  free  and  low frequency  suspension
controllers.  Consequently,  any damage to the  entire experiment when
the  coils are  excited can  be safely  discarded.  Moreover,  we have
shown  as well  that  the  displacement and  rotation  signals can  be
processed  and  pipelined to  an  adequate  estimation algorithm  that
allows  to  estimate  both   the  magnetic  moment  and  the  magnetic
susceptibility of  the test masses to a  good accuracy.  Specifically,
assuming  that the remnant  magnetic moment  is homogenous  within the
entire volume of the test  masses, the estimates have errors below the
2\% level.

\section*{Acknowledgments}
This work  was partially supported by MICINN  grants ESP2007-61712 and
AYA08--04211--C02--01.  Part  of this work  was also supported  by the
AGAUR.

\end{document}